\documentclass[prl,twocolumn,preprintnumbers,nofootinbib,showpacs,amsmath,amssymb,superscriptaddress]{revtex4-1}

\usepackage{color}

\usepackage[
	  pagebackref=false,
	  colorlinks=true,
      linkcolor=blue,
      urlcolor=blue,
      filecolor=black,
      citecolor=blue,
      pdfstartview=FitV,
      pdftitle={},
        pdfauthor={},
        pdfsubject={},
        pdfkeywords={},
        pdfpagemode=None,
        bookmarksopen=true
      ]{hyperref}

\usepackage{amsfonts}
\usepackage{amsmath}
\usepackage{amssymb}
\usepackage{upgreek}
\usepackage{bm}
\usepackage{dcolumn}
\usepackage{epsfig}
\usepackage{graphicx}
\usepackage{graphics}
\usepackage[latin1]{inputenc}
\usepackage{latexsym}
\usepackage{rotating}
\usepackage{tipa}
\usepackage{hyperref}
\usepackage[all]{hypcap}
\usepackage{xspace} 

\definecolor {darkgreen}{rgb}{0.2,0.7,0.2}

\newcommand\be{\begin{equation}}
\newcommand\ba{\begin{eqnarray}}
\newcommand\ee{\end{equation}}
\newcommand\ea{\end{eqnarray}}
\newcommand\bw{\begin{widetext}}
\newcommand\ew{\end{widetext}}
\newcommand{\nn}{\nonumber}
\newcommand{\pd}{\partial}

\newcommand\SR{\, ^*\hspace{-0.07cm}R}

\newcommand{\mrm}{\mathrm}
\newcommand{\lb}{\left(}
\newcommand{\rb}{\right)}

\newcommand\aCS{\alpha_\mathrm{CS}}

\newcommand\al{{\alpha}}
\newcommand\ep{\epsilon}

\newcommand\vt{\vartheta}

\newcommand\de{{\ensuremath{{\delta}}}}

\newcommand\nab{{\nabla}}

\def\th{{\theta}}

\newcommand\ov{\over}
\newcommand\ha{{1 \ov 2}}

\def\le{\left}
\def\ri{\right}

\newcommand\p{\ensuremath{\partial}}

\newcommand\vev[1]{{\ensuremath{\left\langle{#1}\right\rangle}}}

\newcommand\sO{{\ensuremath{{\mathcal O}}}}

\newcommand\sS{{\mathcal S}}

\newcommand\bCS{\beta_\mathrm{CS}}
\newcommand\SF{\, ^*\hspace{-0.07cm}F}

\begin{document}

\preprint{CALT-68-2901, IPMU12-0227, MIT-CTP 4426}

\title{Spontaneous Generation of Angular Momentum in Holographic Theories}

\author{Hong Liu}
\affiliation{Center for Theoretical Physics,
Massachusetts Institute of Technology,
Cambridge, MA 02139, USA}
\author{Hirosi Ooguri}
\affiliation{California Institute of Technology, 452-48, Pasadena, CA 91125, USA}
\affiliation{Kavli Institute for the Physics and Mathematics of the Universe (WPI),
University of Tokyo, Kashiwa 277-8583, Japan}
\author{Bogdan Stoica}
\affiliation{California Institute of Technology, 452-48, Pasadena, CA 91125, USA}
\author{Nicol\'as Yunes}
\affiliation{Department of Physics, Montana State University, Bozeman, MT 59717, USA}

\date{\today}

\begin{abstract} 

The Schwarzschild black two-brane in four-dimensional anti--de Sitter space is dual to a finite temperature state
in three-dimensional conformal field theory. We show that the solution acquires 
a nonzero angular momentum density when a gravitational Chern-Simons coupling is 
turned on in the bulk, even though the solution is not modified. A similar phenomenon is found for
the Reissner-Nordstr\"om black two-brane with axionic coupling to the gauge field. 
We discuss interpretation
of this phenomenon from the point of view of the boundary three-dimensional conformal field theory. 

\end{abstract}


\maketitle


\noindent
{\emph{Introduction.}}--- The gauge-gravity correspondence has provided many important insights into strongly coupled gauge theories.
In particular, parity violating interactions in the bulk have been shown to
generate interesting effects on boundary field theories.  One example is the effect of anomalies 
in four dimensions \cite{anomalyone, anomalytwo, anomalythree} (see also Chap. 20 of \cite{Volovik} and references therein), which had been overlooked in the traditional
approach to hydrodynamics. Another is the existence  
of spatially modulated phase transitions in three and four dimensions \cite{modulatedone, modulatedtwo, modulatedthree, modulatedfour}. 
In this Letter, we point out yet another striking effect of a parity violating interaction --- spontaneous generation of an angular momentum
density and an edge current. This question was previously examined by Saremi~\cite{saremi}.
Parity violation effects in hydrodynamics have been discussed in~\cite{Jensen:2011xb}, which also pointed out angular momentum generation,
though its physical mechanism and its connection to the edge current have not been examined.


The spontaneous generation of angular momentum and an edge current are typical phenomena in parity-violating physics. They occur, for example, in the $A$ phase of helium-3, where the chiral $p$-wave condensate breaks parity~(see, for example,~\cite{Volovik,stone,sauls}). 
There has been a controversy on its value in a given container geometry since different methods give different answers. 
The holographic mechanism to generate the angular momentum 
density described here may provide a new perspective on such macroscopic parity-violating effects. 

We consider here a $(2+1)$-dimensional boundary field theory with a $U(1)$ global symmetry, which is described by classical gravity (together with various matter fields) in a four-dimensional, asymptotically anti--de Sitter spacetime~(AdS$_4$).  The conserved,
$U(1)$ boundary current $j^\mu$  is mapped to a bulk gauge field $A_a$. We use $a, b=0, 1, 2, z$ to denote bulk indices, $\mu,\nu=0,1,2$ for boundary indices and $i, j =1,2$ for boundary spatial indices. 

We discuss two representative bulk mechanisms for the spontaneous generation of angular momentum, with a gravitational Chern-Simons interaction $\int \vartheta\ R \wedge R$~\cite{CSgravity}
and with an axionic coupling $\int \vartheta \ F \wedge F$~\cite{Wilczek:1987, Carroll:1989vb}, 
where $\vartheta$ is a dynamical massless pseudoscalar, which is dual to a marginal pseudoscalar operator $\sO$ 
in the boundary field theory, and $R$ and $F$ are the Riemann curvature two-form and the
field strength for a gauge field $A_a$, respectively. 
To break the parity symmetry, we turn on a non-normalizable mode for the pseudoscalar field $\vartheta$.
With the gravitational Chern-Simons interaction, we obtain a non-zero angular momentum density
at finite temperature. Similarly, the axionic coupling can generate a non-zero angular momentum density
at a finite chemical potential.
In both situations if we put the system in a finite box (i.e., $\vt$ is nonzero only inside the box), 
the spontaneous generation of angular momentum is always accompanied by an edge current.

Without going into details of the bulk calculation, both bulk mechanisms can be understood from the boundary perspective as follows. The constant value $\th$ of the massless pseudoscalar $\vartheta$ is a non-normalizable mode, corresponding to turning on a marginal deformation  $ \th  \int d^3 x \, \sO$ in the boundary theory that breaks parity. The presence of bulk interactions ($\int \vartheta\ R \wedge R$ or $\int \vartheta \ F \wedge F$) generates a mixed two-point function 
\be \label{mixco}
\vev{T_{0i} (x) \sO (y)}_{\th} = - C\ep_{ij} \p_j^{(y)} \de^{(3)} (x-y) + \cdots \ ,
\ee
at a finite temperature or a finite charge density, where  $\ep_{12} = -\ep_{21} =1$, $C$ is a constant depending on the temperature or charge density of the system, and $\cdots$ denotes higher-order derivative terms which are irrelevant here.
 
Now, consider making $\th$ slightly non-homogeneous; then, from~\eqref{mixco} and to leading order in the derivative expansion of $\th$, we have
\be \label{strex}
\vev{T_{0i}}_{\th} = C \ep_{ij} \p_j \th (x) + \cdots 
\ee
which vanishes for constant $\th$. Let us consider a profile of $\th (x)$ which takes constant value $\th_0$ inside a spherical box of size $L$ but eventually goes to zero outside the box along the radial direction. (We use a spherical box for convenience of illustration. Our conclusions do not depend on the shape of the box, as far as it is sufficiently big.) At the end of the calculation we take $L$ to infinity. From~\eqref{strex}, we then find that the angular momentum $J$ of the boundary is given by
\be \label{fabg1}
J =  \ep_{ij} \int d^2 x \,x_i \vev{T_{0j}}_\th= - 2 \; C \; \th_0 \; \int d^2 x \, 
\ee
which remains nonzero for a constant $\th$.  For a finite (but large) $L$,  $\vev{T_{0i}}_\th$ is zero both inside and outside the box, but will be nonzero in the transition region where $\th (x)$ changes from $\th_0$ to zero.  
In other words, there is an edge momentum flow. In terms of the polar coordinate $(r, \phi)$, the nonvanishing component of this edge current is 
\be \label{edcu}
\vev{T_{0 \phi}}_\th = C  h_L (r) + \cdots
\ee
where $h_{L} (r)$ is a function with compact support near $r=L$, and whose precise form depends on the specific  profile of $\th (x)$.  

Heuristically,  $\theta$ can be considered as a measure of the strength of parity breaking. A constant nonzero $\theta$ inside the box has thus two effects: (i) a nonzero angular momentum inside the box; (ii) an edge current at the boundary of the box.

When the system is at a finite charge density, then there is also a parallel story 
for the $U(1)$ charge current $j_i$, with $T_{0i}$ in~\eqref{mixco} and~\eqref{strex} replaced by $j_i$ and $C$ replaced by some other constant $C_{\rm charge}$. We can also define a ``charge angular momentum'' $J_{\rm charge} = \int d^2 x \, \ep_{ij} x_i j_j$. A nonzero $\theta$ inside the box then also leads to  a nonzero charge angular momentum $J_{\rm charge}$ and an edge $U(1)$ current $j_\phi$, which can be obtained by replacing 
$C$ in~\eqref{fabg1} and~\eqref{edcu} by $C_{\rm charge}$.

We now provide an explicit derivation of~\eqref{mixco} and the corresponding $C$ and $C_{\rm charge}$ from bulk gravity. 


{\emph{Gravitational Chern-Simons interaction.}}--- Consider the following action~\cite{Alexander:2009tp}
\be\label{act1}
S = \frac{1}{2\kappa^2}\intop d^4x\sqrt{-g} \left[ R + \frac{6}{\ell^2} - \frac{1}{2}\left(\partial \vartheta\right)^2  -
\frac{\aCS \ell^2}{4}\vartheta \SR R\right]
\ee
where $ \SR R = \SR^{abcd} R_{bacd}$ and $\SR^{abcd} = \ha \ep^{cdef} R^{ab}{_{ef}}$.
 $\ep^{abcd}$ is the totally antisymmetric tensor with  
 $\epsilon^{012z}={1/\sqrt{-g}}$.
The equations of motion are
\ba
\label{eom1}
& & R_{ab} + \frac{3}{\ell^2} g_{ab} = {\aCS}{\ell^2} C_{ab} +\frac{1}{2}\partial_a \vartheta \partial_b \vartheta, \\
\label{eom2}
& & \frac{1}{\sqrt{-g}}\partial_a\left(g^{ab}\sqrt{-g}\partial_b \vartheta \right) =  \frac{\aCS \ell^2}{4} \SR R,
\ea
where $C^{ab} \equiv \nab_c (\nab_d \vt \SR^{c(ab)d})$ and parenthesis in index lists denote symmetrization.  
Equations~\eqref{eom1}--\eqref{eom2} are solved by the standard Schwarzschild black brane
\be
\label{schw}
ds_0^2 = \frac{\ell^2}{z^2}\left[-f(z) dt^2 + \frac{dz^2}{f(z)} + \gamma_{ij} dx^i dx^j\right]\,,
\ee
if $\vt$ is a constant, where $\gamma_{ij}$ is the flat metric in $(x,y)$ space and $f(z) = 1 - z^3/z_0^3$. 
The horizon is located at $z=z_0$ with a temperature $T = 3/(4 \pi z_0)$.

Let us now take the boundary value for $\vt$  to be spacetime dependent $\th (x^\mu)$. Clearly, $\vt (z,x^\mu) = \th (x^\mu)$ and~\eqref{schw} no longer solves~\eqref{eom1}--\eqref{eom2}. Nevertheless, if 
$\th (x^\mu)$ varies slowly  over spacetime, we can solve the bulk equations of motion order by order in a derivative expansion of $\th (x^\mu)$.  In particular, from the modification of the bulk metric, we could read the response of the boundary stress-energy tensor to a nonuniform $\th (x^\mu)$. The calculation is similar in spirit to that of forced fluid dynamics~\cite{hydrofour}, but at the end of the calculation we will take $\th (x)$ to be a constant. For our purpose, it is enough to work out the expansion to first-order in $\p_i \th$ with $\th$ time-independent, in which case only the $g_{0i}$ components of the metric and $\vt$ are modified. To carry out the derivative expansion, it is convenient to introduce the book-keeping parameter $\ep$ to count the number of boundary spatial derivatives, with $\p_i \vt = {\cal{O}}(\ep), \p_i \p_j \vt = {\cal{O}}(\ep^2), \p_i \vt \p_j \vt = {\cal{O}}(\ep^2)$ and so on. 

Writing the metric as
\be \label{mede}
ds^2 = ds_0^2 + 2 {\ell^2 \ov z^2} a_i dx^i dt
\ee
with $(a_{1},a_{2})$ functions of $(z,x,y)$, 
the nontrivial components of the Einstein equations~\eqref{eom1} are the $(z,t)$ component
 \be
\label{div_a}
\partial_i a_i = f(z)G(x^i),
\ee
with $G(x^i)$ an arbitrary function of $x^i$, and the $(t,i)$ components
\be \label{ti}
\ep_{ij} \p_j B - {f \ov z} (z \p_z^2 a_i - 2 \p_z a_i) = - \ep_{ij} {\al_{\rm CS} z f f'' \ov 2} (\p_j \vt + z \p_z \p_j \vt)
\ee 
where $B \equiv \partial_x a_{y} - \partial_y a_{x}$ and $f' = \partial_z f$, etc. Equation~\eqref{eom2} gives (to first order in $\p_i \th$)
\be \label{theq}
z^2 \p_z (z^{-2} f \p_z \vt) =  {\al_{\rm CS} \ov 2} { z^2 f'' } \p_z B \ .
\ee

Since we are considering a normalizable solution for the metric, $G$ must vanish. We thus have $\p_i a_i =0$, which implies 
that $\ep_{ij} \p_j B = - (\p_x^2 + \p_y^2) a_i$. Assuming regularity of $a_i$ and $\vt$ at the horizon, then Eq.~\eqref{ti} implies that $(\p_x^2 + \p_y^2) a_i (z_0, x^i) =0$ at the horizon. Imposing the boundary condition $a_i (z_0, x^i) \to 0$ at spatial infinity $r \to \infty$ [note that this boundary condition is consistent with that for $\th (x)$ as discussed in the paragraph following Eq.~\eqref{strex}], we then conclude that
\be \label{horibd}
a_i (z_0, x^i) = 0
\ee
at the horizon. From Eq.~\eqref{ti}, $a_i \sim {\cal{O}}(\ep)$ and thus $ \p_j B \sim {\cal{O}}(\ep^2)$; i.e., we keep $\ep_{ij} \p_j B$
above only to impose the boundary condition~\eqref{horibd}. Applying $\p_i$ on both sides of~\eqref{theq}, imposing regularity of $\vartheta$ at the horizon,
and keeping terms only to ${\cal{O}}(\ep)$, we find that
\be \label{conth}
\p_z \p_i \vt = 0  \quad \to \quad \p_i \vt (z, x^\mu)= \p_i \th (x^\mu);
\ee
i.e., $\p_i \vt$ is $z$-independent. Now Eq.~\eqref{ti} can be immediately integrated at ${\cal{O}}(\ep)$ to give
\be
\label{ai_is}
a_i = \epsilon_{ij} \frac{3 \aCS z^3 (z_0-z) \partial_j \th}{4z_0^3 } \ ,
\ee
fixed uniquely by normalizability at infinity and~\eqref{horibd}.

We now proceed to compute the boundary stress-energy tensor due to~\eqref{ai_is}. There is an important complication here  as in addition to the standard contribution, there are potential contributions from: (i) direct variation 
of the $\SR R$ term; (ii) additional boundary counter terms required due to the presence  $\SR R$. We discuss these contributions in detail in the Appendix and show that they  
vanish. Therefore, it suffices to use the standard formulas as in~\cite{KB1999, BFS2002, dHSS2001}, which 
give
\be \label{stesr}
T_{0i} ={\ell^2 \ov 2 \kappa^2} \frac{9\aCS \epsilon_{ij} \pd_j \theta}{4 z_0^2} \ .
\ee
Equation~\eqref{stesr} leads to~\eqref{mixco} with
\be \label{hity}
C = {\ell^2 \ov 2 \kappa^2} \frac{9\aCS }{4 z_0^2} =  { \aCS  \ov 2} \sS_3 T^2 =  \frac{9\aCS}{16 \pi}  s.
\ee
Here, $\sS_3 = (2 \pi)^2 \ell^2 / \kappa^2$ is the central charge of the CFT defined either using entanglement entropy on a disk~\cite{Myers:2010xs} or equivalently the free energy on an $S^3$~\cite{Jafferis:2011zi}. Moreover, $s = 2 \pi \ell^2 / \kappa^2 z_0^2$ is the entropy density of the finite temperature system. 

{\emph{Axionic coupling.}}---
Let us now set $\aCS =0$ in~\eqref{act1} and  add to this equation the following terms:
\be
S_{\rm ax} = - \frac{\ell^2}{2\kappa^2}\intop d^4x\sqrt{-g} \left[ F^{ab} F_{ab}   + \bCS \vartheta \SF^{ab} F_{ab} \right],
\ee
with $\bCS$ a dimensionless constant and  $\SF^{ab} \equiv \frac{1}{2}\epsilon^{abcd}F_{cd}$. The equations of motion are now
\ba
\label{eom12}
& & R_{ab} + \frac{3}{\ell^2} g_{ab} - 2\ell^2\left(F_{ca}F^c_{\ b} - \frac{g_{ab}}{4}F^2\right) = \frac{1}{2}\partial_a \vartheta \partial_b \vartheta,\quad  \\
\label{eom11}
& & \frac{1}{\sqrt{-g}}\partial_a\left(g^{ab}\sqrt{-g}\partial_b \vartheta \right) = {\bCS}{\ell^2} \SF F, \\
\label{eom13}
& & \partial_a \left[ \sqrt{-g} \left( F^{a b} + \bCS  \vartheta \SF^{ab}\right)\right] = 0,
\ea
which admit as a solution the standard AdS charged brane if $\vt$ is a constant. The metric has the form~\eqref{schw} but with
\be
f(z) = 1 - \frac{z^3}{z_M^3} + \frac{z^4}{z_Q^4},
\ee
and the gauge potential is
\be
A_t^{(0)} = \mu \le(1 - {z \ov z_0} \ri) , \quad \mu = {z_0 \ov z_Q^2},
\ee
where $z_0$ is the location of the horizon and $\mu$ the chemical potential. 

As before, we take the boundary source $\th (x^i)$ to be spatially inhomogeneous, but slowly varying.  In addition to a metric deformation as in~\eqref{mede}, such a boundary source will now also excite the bulk gauge field $A_i$  along the boundary spatial direction. The analysis of the equations is similar to the previous example; in particular,
the scalar equation still yields~\eqref{conth}, and~\eqref{horibd} also applies. To ${\cal{O}}(\ep)$, the nontrivial equations from~\eqref{eom11} and \eqref{eom13} are
\ba \label{k2}
&&\p_z \le(f z_Q^2 A_i' - a_i \ri) = \bCS \ep_{ij} \p_j \vt  \ , \\
\label{k3}
&& z a_i'' - 2 a_i' - {4z^3 \ov z_Q^2} A_i' = 0 \ ,
\ea
which can be integrated exactly. Upon imposing the normalizability condition at infinity and 
the boundary condition~\eqref{horibd} at the horizon, we find that $a_i$ and $A_i$ have the following leading-order behavior near the boundary:
\ba 
a_i (z) &= & {2 z^3 z_0^2 \ov  3 z_Q^4 } \bCS \ep_{ij} \p_j \th  + {\cal{O}}(z^4) , \\
A_i &=&  -{ z_0 z \ov z_Q^2}  \bCS \ep_{ij} \p_j \th + {\cal{O}}(z^2)  \ .
\ea
We then find the stress-energy tensor and the charged current
\ba \label{stre2}
T_{0i} &=& {\ell^2 \ov 2 \kappa^2} {2  z_0^2 \ov  z_Q^4 } \bCS \ep_{ij} \p_j \th   \ , \\
j_i &=& {4 \ell^2 \ov 2 \kappa^2} { z_0 \ov  z_Q^2}  \bCS \ep_{ij} \p_j \th \ ,
\label{chcu}
\ea
which lead to
\ba
\label{fiep}
C &=&  {\ell^2 \bCS \ov  \kappa^2} { z_0^2 \ov   z_Q^4 }  = \bCS {\pi \rho^2 \ov 2 s}, \\
C_{\rm charge} &=&   {2 \ell^2 \bCS \ov  \kappa^2} { z_0 \ov  z_Q^2}   = {\bCS  \ov 2 \pi^2} \sS_3 \mu,
\ea
where $\rho = 2 \ell^2 / \kappa^2 z_Q^2$ is the charge density, $s=2 \pi \ell^2 / \kappa^2 z_0^2$ is the entropy density, and $\sS_3$ is the central charge as discussed earlier. Note that $C_{\rm charge}$ is temperature independent. In the extremal limit, $s = \lb \pi / \sqrt{3} \rb \rho$, and
we then find that [in the strict extremal limit, the intermediate steps appropriate for a nonzero temperature no longer apply due to singular nature of the extremal horizon, but expression~\eqref{fiep} has a well defined zero temperature limit]
\be 
C = {\sqrt{3} \ov 2}  \bCS \rho, \qquad T=0 \ .
\ee

Finally we can turn on a nonzero $\aCS$ in the charged black brane background (setting $\bCS=0$), so that the corresponding action is
\ba
\label{actrn}
S &=& \frac{1}{2\kappa^2}\intop d^4x\sqrt{-g} \bigg[ R + \frac{6}{\ell^2} - \frac{1}{2}\left(\partial \vartheta\right)^2 \\
 &-& \ell^2 F^{ab} F_{ab} - \frac{\aCS \ell^2}{4}\vartheta \SR R\bigg] \nn
\ea 
and we find that the corresponding $C$ and $C_\mrm{charge}$ are  
\ba 
\label{Crn}
& & C  =  { \aCS  \ov 2} \sS_3 T^2,\\
\label{Cchrn}
& &C_\mrm{charge} = 0.
\ea
$C$ is the same as in the Schwarzschild case and it increases monotonically with temperature, so that in the extremal limit $C=0$. In fact, the result \eqref{Crn} -- \eqref{Cchrn} holds for any black hole solution of action \eqref{actrn} for marginal field $\vartheta$, and it likely also generalizes to more complicated systems as long as the scalar field coupling to $\SR R$ remains marginal. We are currently investigating this direction.


{\emph{Hall viscosity}}.--- Another interesting parity odd response to gravitational perturbations is Hall viscosity, which occurs in quantum Hall states \cite{Hallviscosone}, where it is shown to be proportional to angular momentum density in various examples \cite{Hallviscostwo, Hallviscosthree}.  A holographic model exhibiting Hall viscosity has been proposed \cite{SaremiSon}: the Einstein-scalar system studied in this Letter plus a potential for the scalar field. There, the Hall viscosity coefficient
is shown to be proportional to the normal derivative of the scalar field at the horizon of the black brane. Explicit models with nonzero Hall viscosity have been constructed in \cite{Chen:2012ti, Chen:2011fs}. We have verified the Saremi-Son formula of~\cite{SaremiSon} in our gravitational Chern-Simons setups, but the Hall viscosity turns out to be zero since the scalar field is constant in our solution. 
It should be noted, however, that the holographic model used here is dual to a conformal field theory at finite temperature
and not to a gapped zero temperature state. We hope to investigate the Hall viscosity phenomenon in a more realistic setup in the future.

To summarize, in this Letter we identified from two classes of
holographic models a field theoretical mechanism for spontaneous
generation of a nonzero angular momentum density and edge current.
Although our analysis was restricted to a marginal operator, likely it
is more general, applicable to relevant operators or in the absence of
an external source. We will leave these issues for future investigation.

{\emph{Acknowledgments}.--- We thank J.~Alicea, R.~Loganayagam, L.~Motrunich, M.~Oshikawa, N.~Read, D.~T.~Son, 
and G.~Volovik for discussion and, in particular, O.~Saremi for sharing with us his previous unpublished work. 
H.O. is thankful for the hospitality of the Aspen Center for Physics (NSF Grant No.
1066293). The work of H.O. and B.S. is supported in part by U.S. DOE grant
DE-FG03-92-ER40701. The work of H.O. is also supported in part by
a Simons Investigator grant from the Simons Foundation, JSPS Grant-in-Aid for Scientific Research 
C-23540285, and the WPI Initiative of MEXT of Japan. 
H.L. is thankful for the hospitality of Caltech. The work of H.L. is partially supported by a Simons Fellowship and by the
U.S. Department of Energy (DOE) under cooperative research agreement DE-FG0205ER41360. The work of N.Y. is partially supported by NSF Grant No. PHY-1114374 and NASA Grant No. NNX11AI49G, under 00001944.


\appendix 

\section{Appendix: Boundary Stress-Energy Tensor}

We now turn our attention to the boundary stress-energy tensor computation. A priori, there are 4 possible contributions that need to be accounted for: the usual Gibbons-Hawking boundary term, a term arising from the variation of $S_{CS}$, possible additional terms that must be added for the Dirichlet boundary-value problem to be well-defined and local counterterms. Thus, we can write
\be
T^\mathrm{bdy}_{\alpha \beta} = \frac{1}{2\kappa^2}\left(2K_{\alpha \beta} - 2h_{\alpha \beta}K + T_{\alpha \beta}^\mathrm{cs} + T_{\alpha \beta}^\mathrm{reg} - T_{\alpha \beta}^\mathrm{ct} \right).
\ee
The CFT stress-energy tensor can thus be obtained by computing the boundary stress-energy tensor on a plane at finite $z$ parallel to the boundary, multiplying by an appropriate power of $z$ and taking the $z\rightarrow 0$ limit, according to the standard AdS/CFT dictionary (see for e.g. \cite{KB1999, BFS2002, dHSS2001}).

Let us first concentrate on the CS contribution to the boundary stress-energy tensor. With $\lambda \equiv \aCS\ell^2/{2\kappa^2}$ one can show (see e.g.~\cite{SaremiSon}) that the part of $\delta S_{CS}$ that contributes to $T_{\alpha \beta}^\mathrm{cs}$ is
\ba
\delta S_{CS} = &-&2\lambda  \intop d^4x \sqrt{-g}\nabla_c\left( \vartheta\SR^{b\ cd}_{\ a} \delta\Gamma^a_{\ bd}\right)\nonumber\\\nonumber &+&2\lambda  \intop d^4x \sqrt{-g}\nabla_b\left[\delta g_{ed}\nabla_c\left(\vartheta\SR^{becd}\right) \right]. \\
\ea
Applying Stokes' theorem gives
\ba
\label{deltaSCS}
\delta S_{CS} = &+& 2\lambda  \intop_{\mathcal{\partial M}} d^3x \sqrt{-h} \nabla_\alpha\left(n_c \vartheta\SR^{\alpha acb} \right)  \delta g_{ab}\nonumber \\
&-& 2\lambda  \intop_{\mathcal{\partial M}} d^3x \sqrt{-h} n_c \vartheta\SR^{zacb} \nabla_z \delta g_{ab} \nonumber \\
&+& 2\lambda \intop_{\mathcal{\partial M}} d^3x \sqrt{-h}n_b\nabla_c\left(\vartheta\SR^{becd}\right) \delta g_{ed},
\ea
where $n^{a}$ is the unit normal to constant $z$ surfaces and the subscript $z$ refers to the holographic direction. The first and third terms in Eq.~\eqref{deltaSCS} can in principle contribute to the boundary stress-energy tensor, while the second term appears to be of the type that needs to be regularized by the addition of $T_{\alpha\beta}^\mathrm{reg}$ (analogous to the usual Gibbons--Hawking term). 

One can show that, for any asymptotically AdS space, the second term in Eq.~\eqref{deltaSCS} decays one power of $z$ too fast and thus it vanishes at the boundary. For the asymptotic symmetry to be $O(3,2)$, the asymptotic behavior of the metric components must obey (see \cite{HT:1985bg})
\be
g_{ab} = g^{AdS}_{ab} + h_{ab},
\ee
where
\ba
h_{\alpha\beta} &=& \mathcal{O}\left(z^1\right),\\
h_{z\alpha} &=& \mathcal{O}\left(z^4\right),\\
h_{zz} &=& \mathcal{O}\left(z^5\right).
\ea
Taking $\delta g_{ab} = z^{-2} \delta_a^\alpha \delta_b^\beta f_{\alpha\beta}(x^\gamma)+\mathcal{O}(z^{-1})$, we find 
\be
\SR^{zazb} \nabla_z \delta g_{ab} \sim \mathcal{O}\left(z^3\right)
\ee
and $\sqrt{-h} \; n_z \sim \mathcal{O}\left(z^{-2}\right)$. Therefore, the second term in Eq.~\eqref{deltaSCS} does indeed not contribute at the boundary. A similar analysis shows the first and third terms in Eq.~\eqref{deltaSCS} also decay too fast to contribute to the stress-energy tensor. We thus have that $T^{\mathrm{cs}}_{ab} = 0$.

The above result implies that the regularization contribution $T_{ab}^\mathrm{reg}$ should also vanish. This can be checked in the following way. On power-counting grounds, there are two independent parity-breaking terms that could be used to regularize the action (see e.g.~\cite{Grumilleretal}),
\be
\label{pbct1}
\sqrt{-h} \; \vartheta \; n_a \epsilon^{abcd}K_b^{\ e}\nabla_c K_{de}
\ee
and
\be
\label{pbct2}
\sqrt{-h} \; \vartheta \; \epsilon^{\alpha\beta\gamma}\left(\gamma^\lambda_{\ \alpha \mu}\partial_\beta \gamma^\mu_{\ \ \gamma \lambda}+ \frac{2}{3}\gamma^\lambda_{\ \alpha \mu}\gamma^\mu_{\ \ \beta \rho}\gamma^\rho_{\ \gamma \lambda}\right).
\ee
However, the second term is intrinsic to the $z=\mathrm{const.}$ surface (so it cannot be used for regularization) and it can be checked that the first term decays too fast too contribute at the boundary. Thus, there are no terms that can be used to regularize the action.

One can also consider a surface at finite $z$, and ask whether a regularization procedure is needed to ensure that the Dirichlet boundary-value problem is well-defined on this surface. It can be checked that for general metric and scalar field terms, Eq.~\eqref{pbct1} cannot be used to remove the second term in Eq.~\eqref{deltaSCS} without adding other contributions that also need derivatives specified on the boundary, so the Dirichlet boundary-value problem cannot be well-defined on this surface. However, this is not surprising, since $\SR R$ is a higher derivative term and as such it requires more boundary-value data than an usual two-derivative term.

The problem of regularizing a $\vartheta \SR R$ term has been considered before by \cite{Grumilleretal}, who concluded that \eqref{pbct1} should be added to the action for the Dirichlet boundary-value problem to be well-defined. This need not be in contradiction with our result, since \cite{Grumilleretal} considered the problem only in asymptotically flat space.

Let us now focus on possible counterterms. These can be constructed by adding 
\ba
&\sqrt{-h},& \\
&\sqrt{-h}\vartheta,& \\
&\sqrt{-h}\ ^3\!R,& \quad \sqrt{-h} \vartheta^2
\ea
to the action, as well as Eq.~\eqref{pbct2}. However, the $\sqrt{-h} \; \vartheta$ and $\sqrt{-h} \; \vartheta^2$ terms contain divergences not present in the Gibbons--Hawking term, and for any flat boundary, Eq.~\eqref{pbct2} and the Ricci term $\sqrt{-h}\ ^3\!R$ decay too fast to contribute to $T_{ab}^\mathrm{bdy}$. 

Thus, the renormalized action is simply
\be
\label{actcttr}
S_\mathrm{ren} = S - \frac{1}{\kappa^2}\intop d^3x \sqrt{-h}\left(K + \frac{2}{\ell}\right)
\ee
and the renormalized stress-energy tensor is
\be
\label{setenscttr}
T_{\alpha\beta}^\mathrm{bdy} = \frac{1}{\kappa^2}\left(K_{\alpha\beta}-h_{\alpha\beta}K - \frac{2}{\ell}h_{\alpha\beta} \right).
\ee


\begin{thebibliography}{99}
\bibitem{anomalyone} 
  J.~Erdmenger, M.~Haack, M.~Kaminski, and A.~Yarom,
  \href{http://dx.doi.org/10.1088/1126-6708/2009/01/055}{J. High Energy Phys. 01 (2009) 055}.

\bibitem{anomalytwo} 
  N.~Banerjee, J.~Bhattacharya, S.~Bhattacharyya, S.~Dutta, R.~Loganayagam, and P.~Surowka,
  \href{http://dx.doi.org/10.1007/JHEP01(2011)094}{J. High Energy Phys. 01 (2011) 094}.

\bibitem{anomalythree} 
  D.~T.~Son and P.~Surowka,
  \href{http://dx.doi.org/10.1103/PhysRevLett.103.191601}{Phys. Rev. Lett. {\bf103}, 191601 (2009)}.
  
\bibitem{Volovik} 
  G.~E.~Volovik,
  \emph{The Universe in a Helium Droplet} (Oxford
University Press, New York, 2003).

\bibitem{modulatedone} 
  S.~Nakamura, H.~Ooguri, and C.~-S.~Park,
  \href{http://dx.doi.org/10.1103/PhysRevD.81.044018}{Phys. Rev. D {\bf81}, 044018 (2010)}.

\bibitem{modulatedtwo} 
  H.~Ooguri and C.~-S.~Park,
  \href{http://dx.doi.org/10.1103/PhysRevD.82.126001}{Phys. Rev. {\bf D 82}, 126001 (2010)}.

\bibitem{modulatedthree} 
  A.~Donos and J.~P.~Gauntlett,
  \href{http://dx.doi.org/10.1007/JHEP12(2011)091}{J. High Energy Phys. 12 (2011) 091}.

\bibitem{modulatedfour} 
  O.~Bergman, N.~Jokela, G.~Lifschytz, and M.~Lippert,
  \href{http://dx.doi.org/10.1007/JHEP10(2011)034}{J. High Energy Phys. 10 (2011) 034}.

\bibitem{saremi}
O.~Saremi, ``Hall viscosity, angular momentum
and the gravitational Chern-Simons term'' (unpublished). 

\bibitem{Jensen:2011xb} 
  K.~Jensen, M.~Kaminski, P.~Kovtun, R.~Meyer, A.~Ritz, and A.~Yarom,
  \href{http://dx.doi.org/10.1007/JHEP05(2012)102}{J. High Energy Phys. 05 (2012) 102}.

\bibitem{stone} M.~Stone and R.~Roy,
\href{http://dx.doi.org/10.1103/PhysRevB.69.184511}{Phys. Rev. B {\bf69}, 184511 (2004)}.


\bibitem{sauls}   J. A. Sauls,
\href{http://dx.doi.org/10.1103/PhysRevB.84.214509}{Phys. \ Rev. \  B {\bf 84}, 214509 (2011)}.

\bibitem{CSgravity} 
  R.~Jackiw and S.~Y.~Pi,
  \href{http://dx.doi.org/10.1103/PhysRevD.68.104012}{Phys.\ Rev.\ D {\bf 68}, 104012 (2003)}.
  
\bibitem{Wilczek:1987}
  F.~Wilczek,
  \href{http://dx.doi.org/10.1103/PhysRevLett.58.1799}{Phys.\ Rev.\ Lett. {\bf 58}, 1799 (1987)}.  
     
\bibitem{Carroll:1989vb} 
  S.~M.~Carroll, G.~B.~Field, and R.~Jackiw,
  \href{http://dx.doi.org/10.1103/PhysRevD.41.1231}{Phys.\ Rev.\ D {\bf 41}, 1231 (1990)}.

\bibitem{Alexander:2009tp} 
  S.~Alexander and N.~Yunes,
  \href{http://dx.doi.org/10.1016/j.physrep.2009.07.002}{Phys.\ Rep.\  {\bf 480}, 1 (2009)}.


\bibitem{hydrofour}
  S.~Bhattacharyya, R.~Loganayagam, S.~Minwalla, S.~Nampuri, S.~P.~Trivedi, and S.~R.~Wadia,
  \href{http://dx.doi.org/10.1088/1126-6708/2009/02/018}{J. High Energy
Phys. 02 (2009) 018}.



   
   
     
\bibitem{KB1999}
  P.~Kraus and V.~Balasubramanian,
  \href{http://dx.doi.org/10.1007/s002200050764}{Commun.\ Math.\ Phys. {\bf 208}, 413 (1999)}.
  
\bibitem{BFS2002}
  M.~Bianchi, D.~Z.~Freedman, and K.~Skenderis,
  \href{http://dx.doi.org/10.1016/S0550-3213(02)00179-7}{Nucl.\ Phys.\ {\bf B631}, 159 (2002)}.
  
\bibitem{dHSS2001}
  S.~de~Haro, K.~Skenderis, and S.~N.~Solodukhin,
  \href{http://dx.doi.org/10.1007/s002200100381}{Commun.\ Math.\ Phys. {\bf 217}, 595 (2001)}.
 
\bibitem{Myers:2010xs} 
  R.~C.~Myers and A.~Sinha,
  \href{http://dx.doi.org/10.1103/PhysRevD.82.046006}{Phys.\ Rev.\ D {\bf 82}, 046006 (2010)}.
 
\bibitem{Jafferis:2011zi} 
  D.~L.~Jafferis, I.~R.~Klebanov, S.~S.~Pufu, and B.~R.~Safdi,
  \href{http://dx.doi.org/10.1007/JHEP06(2011)102}{J. High Energy Phys. 06 (2011) 102}.

\bibitem{Hallviscosone} 
  J.~E.~Avron, R.~Seiler, and P.~G.~Zograf,
  \href{http://dx.doi.org/10.1103/PhysRevLett.75.697}{Phys.\ Rev.\ Lett.\  {\bf 75}, 697 (1995)}.

\bibitem{Hallviscostwo} 
  N.~Read,
  \href{http://dx.doi.org/10.1103/PhysRevB.79.045308}{Phys.\ Rev.\ B {\bf 79}, 045308 (2009)}.

\bibitem{Hallviscosthree} 
  A.~Nicolis and D.~T.~Son,
  \href{http://arXiv.org/abs/1103.2137}{arXiv:1103.2137}.

\bibitem{SaremiSon} 
  O.~Saremi and D.~T.~Son,
  \href{http://dx.doi.org/10.1007/JHEP04(2012)091}{J. High Energy Phys. 04
(2012) 091}.

\bibitem{Chen:2012ti}
  J.-W.~Chen, S.-H.~Dai, N.-E.~Lee, and D.~Maity,
  \href{http://dx.doi.org/10.1007/JHEP09(2012)096}{J. High
Energy Phys. 09 (2012) 096}.

\bibitem{Chen:2011fs}
  J.-W.~Chen, N.-E.~Lee, D.~Maity, and W.~-Y.~Wen,
  \href{http://dx.doi.org/10.1016/j.physletb.2012.05.026}{Phys.\ Lett.\ B {\bf 713}, 47 (2012)}.
      
\end{thebibliography}

\begin{thebibliography}{99}

\bibitem{HT:1985bg}
  M.~Henneaux and C.~Teitelboim,
  \href{http://dx.doi.org/10.1007/BF01205790}{Commun.\ Math.\ Phys. {\bf 98}, 391 (1985)}.  
  
\bibitem{Grumilleretal}
  D.~Grumiller, R.~Mann and R.~McNees,
  \href{http://dx.doi.org/10.1103/PhysRevD.78.081502}{Phys.\ Rev.\ D {\bf 78}, 081502(R) (2008)}.  
  
\end{thebibliography}
\end{document}